# Generalizable Cone Beam CT Esophagus Segmentation using Physics-Based Data Augmentation

Sadegh R Alam[1], Tianfang Li[1], Pengpeng Zhang[1], Si-Yuan Zhang[2], Saad Nadeem[1]

[1] Department of Medical Physics, Memorial Sloan Kettering Cancer Center, New York, NY, USA

[2] Department of Radiation Oncology, Peking University Cancer Hospital, Beijing, Beijing, CN

## Abstract

**Purpose:** Automated segmentation of esophagus is critical in image guided/adaptive radiotherapy of lung cancer to minimize radiation-induced toxicities such as acute esophagitis. We developed a semantic physics-based data augmentation method for segmenting esophagus in both planning CT (pCT) and cone-beam CT (CBCT) using 3D convolutional neural networks.

**Methods:** 191 cases with their pCT and CBCTs from four independent dataset were used to train a modified 3D-Unet architecture and a multi-objective loss function specifically designed for soft-tissue organs such as esophagus. Scatter artifacts and noises were extracted from week 1 CBCTs using power law adaptive histogram equalization method and induced to the corresponding pCT where reconstructed using CBCT reconstruction parameters. Moreover, we leverage physics-based artifact induction in pCTs to drive the esophagus segmentation in real weekly CBCTs. Segmentations were evaluated using geometric Dice and Hausdorff distance as well as dosimetrically using mean esophagus dose and $D_{5cc}$.

**Results:** Due to the physics-based data augmentation, our model trained just on the synthetic CBCTs was robust and generalizable enough to also produce state-of-the-art results on the pCTs and CBCTs, achieving 0.81 and 0.74 Dice overlap.

**Conclusion:** Our physics-based data augmentation spans the realistic noise/artifact spectrum across patient CBCT/pCT data and can generalize well across modalities, eventually improving the accuracy of treatment setup and response analysis.

# Introduction

Radiotherapy-induced esophagitis is common and leads to dose-limiting acute toxicity in patients with thoracic malignancy, especially when given with concurrent chemotherapy (Alam *et al.*, 2021; Thor *et al.*, 2019). Grade 2 or higher esophagitis could impact the quality of life significantly and lead to disruption or even stop radiotherapy (RT) mid-treatment (Bar-Ad *et al.*, 2014; Fogh *et al.*, 2017). Even though the anatomic delineation of the esophagus has been standardized (Kong *et al.*, 2011), it remains difficult and controversial to define its boundaries on planning CT (pCT) because of esophageal inconsistent appearance and lack of contrast to surrounding tissues (Dieleman *et al.*, 2007). Cone-Beam Computed Tomography (CBCT) is frequently used in clinic for patient setup and treatment response evaluation etc. However, due to artifacts/noise in CBCTs, soft-tissue organs such as esophagus and lymph nodes are difficult to recognize on weekly CBCTs.

In addition, the inter-fraction displacement and deformation of esophagus during the actual treatment have a critical effect on the delivered dose (Abbas *et al.*, 2014; Collier *et al.*, 2003). CBCT images can be used for segmenting tumor/normal tissues e.g. in an adaptive radiotherapy framework where the treatment is adapted mid-way to the weekly changes of the critical organs, such as esophagus, so that the physicians can re-evaluate their plan. To minimize the risk of this treatment related complication, it is necessary to adequately spare esophagus and evaluate accurate dose during the treatment.

Sparing esophagus as the major organ at risk (OAR) during the treatment of lung cancer patients is important due to its proximity to tumor. Due to poor sparing, currently 50% of the patients develop acute esophagitis (Thor *et al.*, 2019; Alam *et al.*, 2021) that severely impacts their quality of life and there is no reliable way to assess the response in real time. The esophagus is a mobile tubular soft tissue organ and the esophagus contour on pCT does not accurately represent the actual weekly esophagus due to anatomical changes and setup uncertainties arising from e.g. physiological variations and respiratory motion over the course of RT (Velec *et al.*, 2011). Discrepancies between the planned and on-treatment esophagus structures are challenging to detect using CBCT due to low soft-tissue contrast (Botros *et al.*, 2015; van Rossum *et al.*, 2015) which is the current limitation of image-guided RT. CBCTs are frequently used in clinic for patient setup and treatment response evaluation, however due to artifacts/noise in CBCTs, soft-tissue organs such as esophagus are difficult to delineate on CBCTs. Being able to properly adopt CBCTs daily/weekly in the clinic, could overcome the image-guided RT limitations.

Few works have been published for auto-segmentation of esophagus using more traditional approaches such as atlas-based (Wang and Jiang, 2004), Skeleton-shape (Feulner *et al.*, 2011), Markov- chain models (Feng *et al.*, 2019). Recently, deep learning models due to their higher generalizability and instance validations were utilized to perform (semi)auto-segmentation of esophagus (Chen *et al.*, 2019; Feng-ming (Spring) Kong *et al.*, 2014; Stark, 2000). Among these studies, some used 2D models to overcome the training dataset size limitation (Cohen *et al.*, 2018; Chen *et al.*, 2019). These however resulted in discontinuity of the generated feature maps which in turn resulted in poor esophagus boundary delineation. Moreover, 2D models require additional data-specific post-processing that makes it less generalizable. This problem is more prominent when segmenting a long tubular soft-tissue organ such as esophagus in noisy CBCT images where 3D spatial information is critical to avoid discontinuities, to minimize the impact of noise/artifacts, and to extract sharper edges with specific tailored network architecture.

Some groups have tried to auto-segment esophagus on high-quality pCT images that are used to delineate tumors and OARs for treatment planning in RT (Dong *et al.*, 2019; Cohen *et al.*, 2018). However, no work has been done on segmenting esophagus on low-quality CBCT images. Delineation of esophagus is challenging even for the physicians on pCT and CBCTs due to low soft-tissue contrast and the presence of artifacts and noise; the best previous Dice overlap achieved by deep learning algorithms on high-quality pCTs was 0.72 (Yang *et al.*, 2018). Other groups have tried to remove noise/artifacts from CBCTs to make it more similar to pCT (Xie *et al.*, 2018; Zhi *et al.*, 2019), however due to the complex nature of these approaches and lack of training data, they only work on individual 2D slices and are not readily usable in clinic.

In this work, we induced different variations of scatter artifacts extracted from the weekly CBCTs to the pCTs for each patient. We then reconstructed the artifact-induced pCTs using CBCT reconstruction parameters/technique. We refer to these artifact-induced pCTs as synthetic CBCTs (sCBCT). Subsequently, sCBCTs were fed to a modified 3D-UNet tailored for esophagus using a multi-objective Dice and binary cross entropy loss function to obtain the segmentation. By adding the noise/scatter artifacts from the low-dose CBCTs to the corresponding pCTs and reconstructing these using a CBCT reconstruction algorithm/parameters, we generate sCBCTs training data that preserves esophagus contours of pCTs. This presents a novel way to train a deep learning model for a semantic and physics-based segmentation of esophagus in both pCTs and CBCTs. In essence, rather than performing a harder task of improving the image quality of CBCTs to match pCTs (as done in previous works), we perform a simpler task of degrading pCTs to better simulate weekly CBCTs (while taking advantage of high-quality pCT contours) which in turn are used for training a deep learning model to segment a focused area of esophagus on real CBCTs. and due to the generalizability of the resultant model on pCTs as well.

## Materials and Method

Fig. 1 shows our entire workflow for generating synthetic CBCTs which was used to train 3D-UNet model. Final model was validated on pCTs and clinical CBCTs from four different types of datasets.

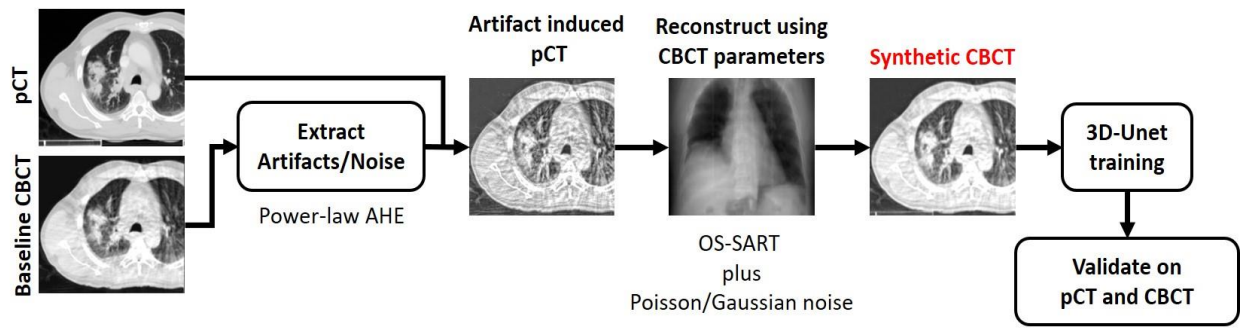

Fig. 1. Entire workflow for generating synthetic CBCTs which in turn are fed to train a 3D-UNet. The model was externally validated using pCT and real weekly CBCTs.

### Dataset

This study included three independent datasets for training and testing/validation and one externally set-aside data:

1) 100 non-small cell lung cancer patients from our institution all treated between June 2016 and March 2019 with intensity-modulated RT in 2 Gy daily fractions in a five days/week fractionation (total dose 60/66/70 Gy) with concurrent chemotherapy (hereafter referred to as MSK cohort). All patients had daily kV orthogonal radiographs for daily setup and weekly CBCTs (matched on spine) to monitor positional uncertainties and tumor changes (Varian Linac TrueBeam; Varian Medical Systems, Palo Alto, CA). At simulation, all patients were immobilized and a free-breathing pCT scan (Philips Big Bore) was acquired for treatment planning (Eclipse; Varian Medical Systems) and a 4DCT to define Internal Target Volume and evaluate tumor motion. The resolution of the pCTs and wCBCTs were $1.17\times1.17\times3$ mm$^3$ and $0.98\times0.98\times3$ mm$^3$, respectively. The esophagus and Gross Tumor Volume (GTV) on the pCTs and weekly CBCTs were contoured (Varian Eclipse) by an experienced radiation oncologist (RO) and represented the ground-truth contour. The esophagus on CBCTs was contoured according to anatomic and contouring atlases of organs at risk (Kong *et al.*, 2011; Feng-ming (Spring) Kong *et al.*, 2014). The esophagus was outlined from the level below the cricoid to the stomach entrance at the GE junction. Altering the CT window/level assisted the visualization and delineation of the boundary of the esophagus in continuous coronal and sagittal views. At the locations where distinguishing the esophageal boundary with adjacent mediastinal lymphadenopathy was challenging, the pCT was used as a reference. To assess the reliability/accuracy of the manual contours, the RO blindly re-delineated the contours for a random subset of 10 patients and the intra-observer delineation variation was reported (R Alam *et al.*, 2020).

2) 13 patients from an external publicly available longitudinal dataset (Hugo *et al.*, 2017) originally collected for investigating breathing patterns of LA-NSCLC patients undergoing 3D conformal radiation therapy. In this public dataset, planning 4DCT and 6 weekly 4D-CBCTs were provided with planning esophagus contours. Image resolution for pCT and weekly CBCT were $0.98\times0.98\times3$ mm$^3$. For this dataset, contours of esophagus were first propagated from pCT to weekly CBCT via Deformable Image Registration (DIR) (Riyahi *et al.*, 2018; Tustison and Avants, 2013; Alam *et al.*, 2020), and a radiation oncologist reviewed and modified (when necessary) the contour on the weekly CBCT serving as the ground truth. (hereafter referred to as external LIDC cohort).

3) 60 patients from public AAPM challenge data (Yang *et al.*, 2018) (hereafter referred to as AAPM cohort). This cohort only included pCT images.

Another externally set-aside cohort from our institution consisted of 18 LA-NSCLC patients with weekly CBCTs treated with the same criteria as the training cohort was also used for completely independent validation (hereafter referred to as MSK validation cohort).

## Generating Synthetic-CBCT images

**Deformable registration**

First, week 1 CBCTs (target) were deformably registered to their pCTs (reference) using a BSpline regularized diffeomorphic image registration (Riyahi *et al.*, 2018; Tustison and Avants, 2013; Alam *et al.*, 2020). Since registration of longitudinal images over the course of RT is challenging due to loss/gain of tumor mass/volumes and lack of correspondence between the two images, the integrated B-spline regularization fits the deformation vector field to a B-spline object to capture large differences. This gives free-form elasticity to the converging/diverging vectors that represents a morphological shrinkage/expansion in tumor and OARs and significantly improves the alignment of corresponding structures.

**Artifact extraction**

Different variations of scatter artifacts/noise were extracted from the registered week 1 CBCTs that contained the highest to the smoothest frequency components using a Signed Power-Law Adaptive Histogram Equalization (PL-AHE) (Stark, 2000). Originally, histogram equalization (HE) is defined as applying a mapping function to the quantized gray-levels of an image $u(x,y)$ and obtain a relatively uniform intensity distribution. To achieve this, we first find the local histogram $\hat{h}(u(x,y))$ by convolving over the image using a kernel window (Stark, 2000). Then, to estimate a mapping that produce an image $u(T(x,y))$ with uniform histogram, a mapping function (T) can be described as:

$$T(x,y) = \sum \hat{h}(u(x,y)) f_c(u,v) \quad (1)$$

Hence, T is a cumulative histogram, characterizing the distribution of gray-levels and $f_c$ is the cumulation function and for PL-AHE is defined as Eq. 2, in which $u - v$ describes subtraction of any gray levels calculated from the cumulative histogram statistics inside the kernel window (Stark, 2000):

$$f_c(u,v) = q(u-v,\alpha) - \beta q(u-v,1) + \beta u \quad (2)$$

where

$$q(u-v,\alpha) = \frac{1}{2} sign(u-v)|2(u-v)|^\alpha \quad (3)$$

$[sign(x): 1 \text{ if } x > 0; 0 \text{ if } x = 0; -1 \text{ if } x < 0]$ and $[0 \leq \alpha, \beta \leq 1]$.

Here, $f_c(u,v)$ is the adaptive transformer of $\hat{h}$ in the process of AHE. Note that in a standard AHE, simply $f_c(u-v) = \frac{1}{2} sign(u-v)$. PL-AHE provides contrast effect hyper-parameters α, β to tune within different variations of AHE. The parameter α controls how similar PL-AHE acts as the standard AHE method (α=0) or as a local-mean subtraction (unsharp masking) high-pass filter (α=1) via a power-law relationship (Eq. 3). Parameter β (beta) on the other hand is a proportion and can be tuned to regulate the output contrast, ranging from the original image to a PL-AHE. Setting (β=0, α≠0) corresponds to unsharp masking the image. However, the final contrast is conceived from tuning a combination of both parameters. For example, by (α=1 & β=0) algorithm acts closer to a high-pass filter hence enhance a dynamic ranges of intensity distribution (artifact/noise) and setting (α=0 & β=1) the algorithm acts more similar to a low-pass filter result in more uniform and smoother output. (β=1 & α=1) results in an almost unmodified image. Therefore, parameters α and β control the degree of frequency for the extracted contrast in an image.

The advantages of PL-AHE over the conventional HE are (1) power-law transformation of each pixel derived from its neighbors optimized by the parameters α and β, instead of applying a strict pixel-size HE. (2) modifying α & β, we could extract details in an image without significantly changing the desired structures. (3) selecting the combination of parameters makes it easier to examine different frequency components from an image along with a fast and flexible implementation.

We used a fixed window radius (5×5×5 pixels) to convolve over the image and experimentally selected 7 different combinations of α and β to be able to cover large range of frequency components. The combinations were (i) α=0.5 & β=1, (ii) α=1 & β=0.5, (iii) α=0.5 & β=0.5, (iv) α=1 & β=0, (v) α=0.5 & β=0, (vi) α=0 & β=1 and (vii) α=0 & β=0.5. We used C++ Insight Toolkit (ITK) libraries (Ibanez *et al.*, 2005) to implement PL-AHE and extract the artifacts.

**Reconstructing and generating synthetic CBCTs**

Compared to pCT, CBCTs had smaller FOV in cranial-caudal direction. Because primary goal of this work was to segment esophagus in clinical CBCT images, we avoided unnecessary slices that are not included in CBCT FOV. Therefore, we first 1) crop the pCT according to the overlapped FOV between the two images. Next, we performed a 2) pixel-wise addition of the cropped pCT and extracted artifact-only images in the overlapped area. After that, the resulting artifact-induced pCT intensities were 3) rescaled to [0,1].

Then, 2D x-ray projections were generated from the artifact-induced pCTs using the 3D texture memory linear interpolation of the integrated sinograms. The projections were reconstructed using iterative Ordered-Subset Simultaneous Algebraic Reconstruction Technique (OS-SART)

(Wang and Jiang, 2004) to produce synthetic CBCTs (sCBCT); Simple Gaussian noise was further added to the projections before reconstruction to better match the quality of sCBCT with the original w1 CBCT.

The geometric parameters used for reconstruction were taken directly from the clinical practice for CBCT reconstruction: Detector size was 512×512, Detector pixel size was 1 mm×1 mm, Object size and resolution were the same as imported cropped pCT, Distance Detector Source was set to 1500 mm, Distance Detector Object was 1000 mm with Center offset=[-160 mm, 0]. 500 projections were generated through 360-degree rotation.

### sCBCT evaluation methods

sCBCTs were quantitatively evaluated against their ground-truth week 1 CBCT by first comparing their histogram distributions and then using four types of similarity metrics i.e. structure similarity index metric (SSIM), root mean square error (RMSE), cross correlation (CC) (Kim *et al.*, 2015)and universal quality index (UQI) (Zhang *et al.*, 2014).

### 3D-UNet model

We trained a 3D-UNet deep learning model (Fig. 2) using sCBCT images. A multi-objective loss function combining Dice coefficient and binary cross-entropy was optimize using ADAM algorithm where the initial and consistent learning rate and maximal number of epochs were set to $10^{-4}$ and 50, respectively.

A convolution neural networks (CNN) extracted salient imaging features from the sCBCTs and they were concatenated with the deconvolved features as a feedback. Typically, in a conventional UNet, convolutions to extract feature maps are performed over the same resolution of images in each stage and the extracted feature maps from the last layers of each encoded convolutions are sent to the decoder phase and concatenated to the deconvolved feature maps. In this work, we kept the same size/resolution of the features until the second layer, however we down-sampled the convolved feature maps at the last layer. Then subsequently, sending the feature map from the second layer with higher resolution/size to the decoder phase for concatenation. The reason for this was as mentioned before, unlike typical other structures such as tumor, heart, liver etc. that are assumed to have a sphere-like shape, esophagus is a long/curved tubed shape structure and covering the entire connected parts when passing the feedback to the decoder phase was important. Therefore, while we down-sampled the final layer of each CNN to extract more details and passed to the next CNN of each encoder, however, we passed the second layer (not the last) to the decoder to make sure all the tube shape structure was involved. Feature maps illustrated in Fig. 2 depict this concept.

In addition to commonly used weight regularization/dropouts to avoid overfitting, using variations of semantic physics-based synthetic CBCTs to further lower the risk of overfitting increased the training size. As a post-processing, after the binary segmentation, morphological filters were used to fill the holes and remove islands.

In addition to deterministic physics-based augmentation, we also applied 8 types of typical geometric-based data augmentation, including Sharpening (emphasis high frequency components), Sigmoid contrast (enhance soft-tissue contrast), Affine (30% scale enlarge, $10^o$ rotate), Affine(30% scale enlarge, $-10^o$ rotate), Affine (20% scale shrink, $10^o$ rotate), Affine(20% scale shrink, $-10^o$ rotate), Affine (20 degree shear), Affine (-20 degree shear) (Jung, 2018).

The sCBCTs were split into 80/20 training and testing cases and fed to a 3D-UNet for esophagus segmentation using pCT esophagus contours as ground-truth. For comparison, in addition to sCBCT model, we also trained the same model with high-quality pCT and week 1

CBCT images (with geometric data augmentation only) to investigate the impact of the semantic physics-based data augmentation. After all the physics-based and geometric augmentations, 2898 images (3D) were generated and incorporated for training sCBCT model, 1134 images for pCT and 810 images for CBCT models. For testing/validation, 329 images, 47 images and 23 images were fed to sCBCT, pCT and CBCT models, respectively (all 3D image).

In addition to the testing cases of each model, the three models were externally validated on the week 1 CBCTs and pCTs using Dice Similarity Coefficient of overlap and Hausdorff distance between the physician-contoured and UNet-segmented esophagus. We also evaluated dosimetric agreement (Mean Esophagus Dose and $D_{5cc}$) between the ground-truth and the segmented contours using Bland-Altman method (Bland and Altman, 2007).

We implemented 3D-Unet using TensorFlow library on a Python platform. Training and testing/validation were performed on an institutional Lilac GPU High-Performance Computer cluster equipped with 72 Ge-force GTX1080 GPUs (8 GB of memory for each GPU).

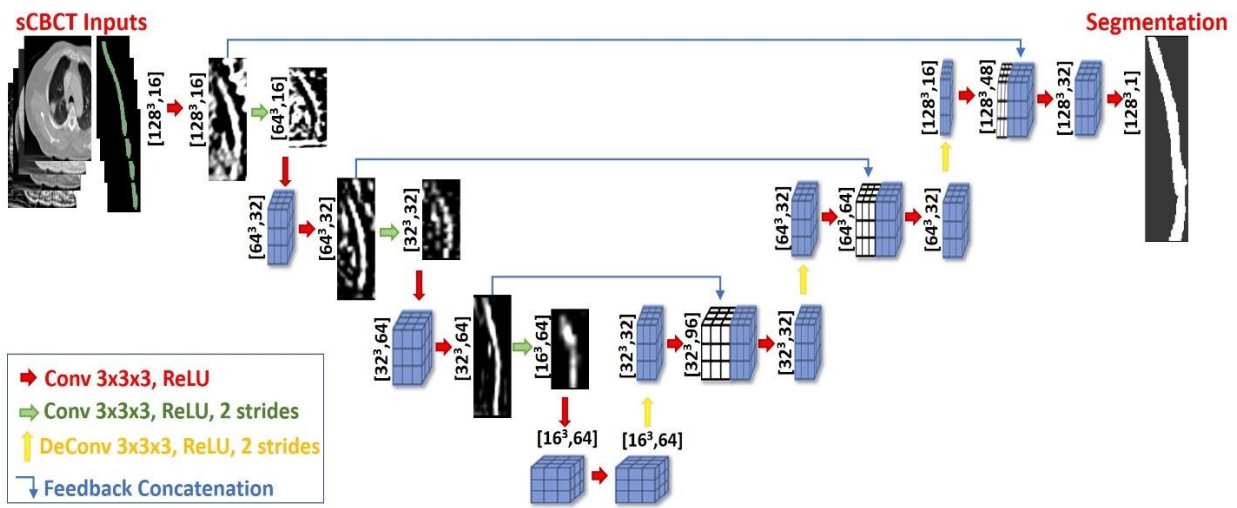

Fig. 2. Proposed 3D-UNet architecture.

# Results

The volumetric change of the esophagus at the end of treatment relative to the start ranged from 40% shrinkage to 82% expansion. Average Dice and maximum (95[th] percentile) HD calculated from the intra-observer variation of contour delineation by the RO on CBCT were 0.89±0.02 and 4.1 mm (1.9 mm) for the first two weeks, respectively.

## Quantitative evaluation of sCBCT images

Different variations of sCBCTs from the highest (top-right, sCBCT 1, SSIM=0.91) to the lowest (bottom-right, sCBCT 5, SSIM=0.76) similarity with the week 1 CBCT are shown in Fig. 3 for a case along with their pCT and ground-truth week 1 CBCTs. These sCBCTs mimic realistic scatter artifacts which could actually be seen in the weekly CBCT acquisitions in clinic.

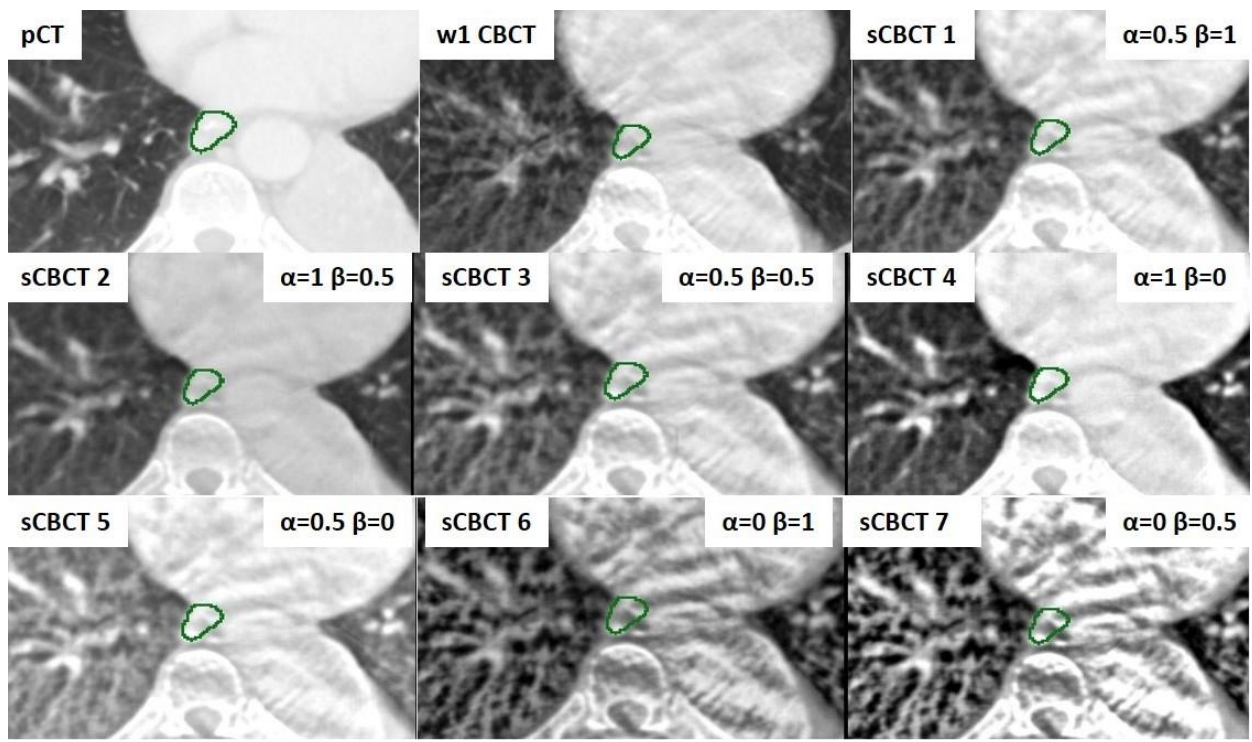

Fig. 3. Example of sCBCT images along with their pCT and the ground-truth week 1 CBCT. sCBCTs are shown (top-right to bottom-right) in the order of the highest to the lowest similarity with the week 1 CBCT. Green contours are ground-truth esophagus contours. Each sCBCT represents different variation of scatter/noise artifact.

The best reconstructed sCBCTs ($\alpha=0.5$, $\beta=1$) had average SSIM=0.89, RMSE=0.05, CC=0.97 and UQI=0.95 in the cohort and the worst ($\alpha=0$, $\beta=0.5$) had SSIM=0.44, RMSE=0.14, CC=0.81 and UQI=0.74 (Fig. 4). Quantitative sCBCT evaluation using four types of similarity metrics showed when $\alpha=0.5$ $\beta=1$, the sCBCT was the most similar to the ground-truth w1 CBCT and when $\alpha=0$ $\beta=0.5$, the least similar (Fig. 4).

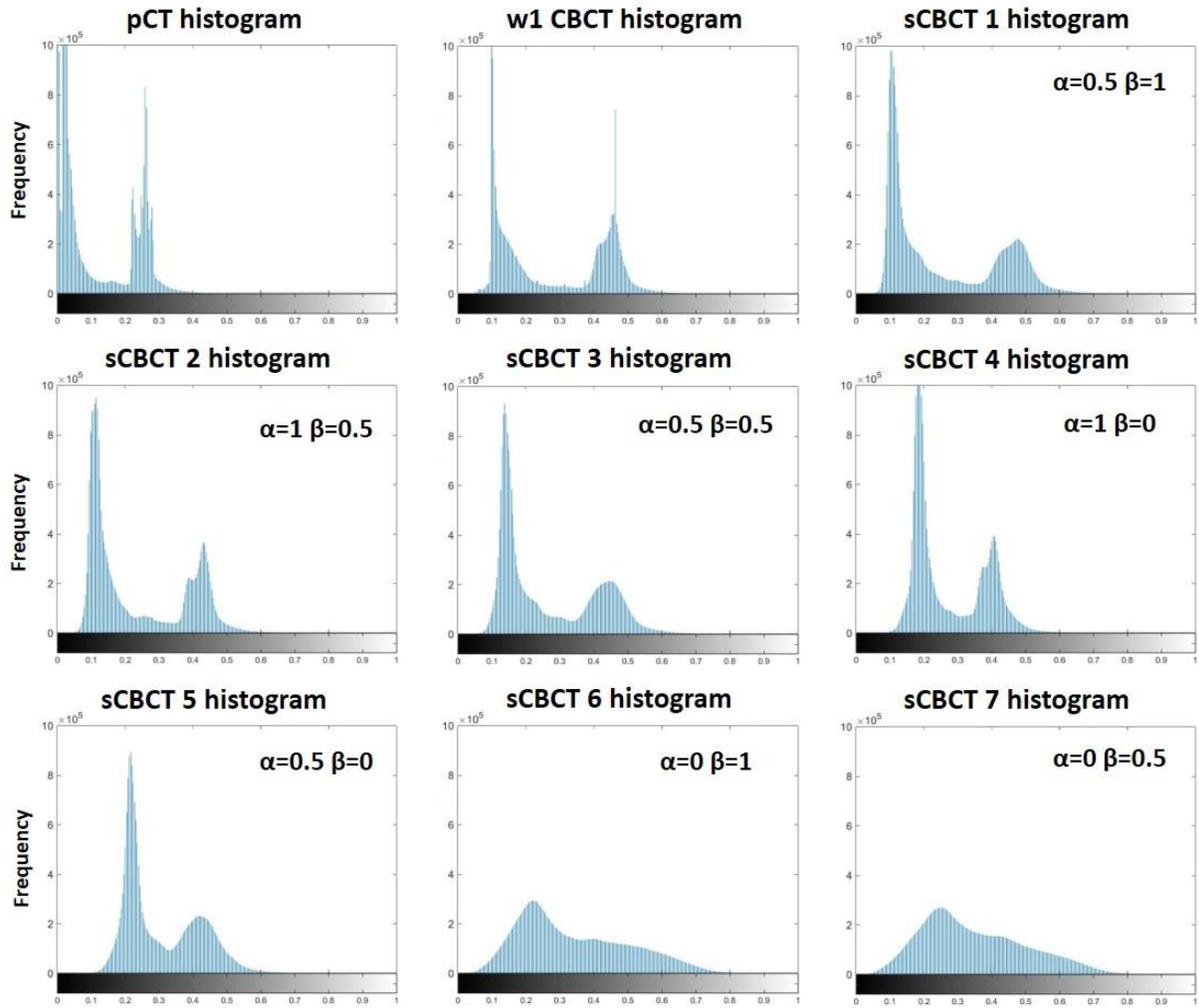
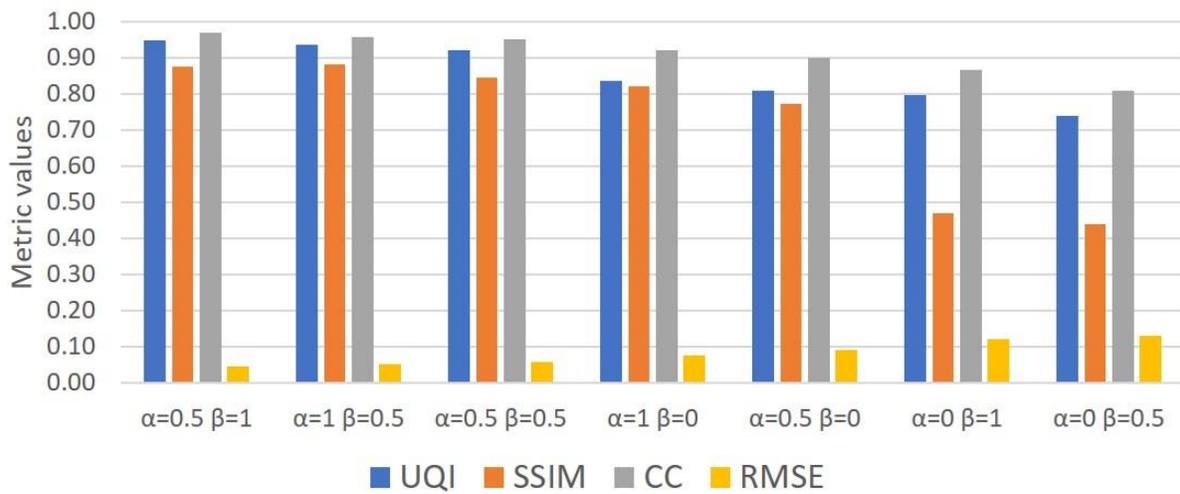

Fig. 4. (top) Histograms of the images presented in Fig. 3 for pCT, week 1 CBCT (ground-truth) and all variations of generated sCBCT. Frequency values in y-axis was normalized to the maximum value in the CBCT images. (bottom) Quantitative sCBCT evaluation (bar graph) using four types of similarity metrics.

Esophagus segmentation

Fig. 5 shows segmented (red) vs. the ground-truth esophagus contours (green) of three typical testing cases, on pCT, ground-truth week 1 CBCTs and their sCBCT with the highest similarity to the ground-truth. The model could segment esophagus on both high quality pCTs as well as low-quality week 1 CBCTs accurately. Scattering artifacts, over exposed high contrast area and also background noise in sCBCTs simulated very similar image quality with their ground truth CBCTs.

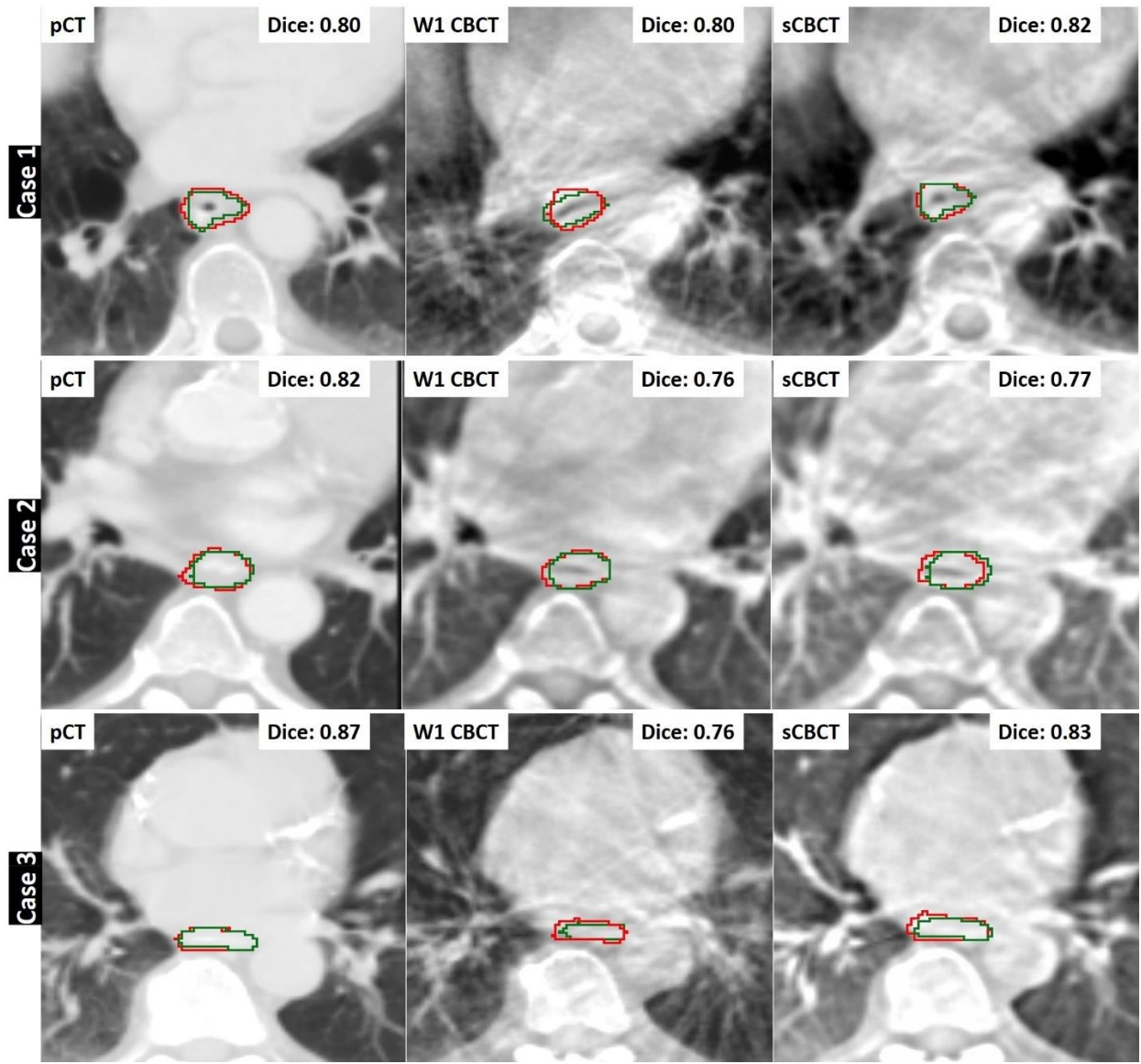

Fig. 5. pCT, week 1 ground-truth CBCTs along with their sCBCT are shown for three typical cases. For each case, green and red contours are the ground-truth and segmented esophagus contours, respectively.

Validation results for all three different models i.e. sCBCT, CBCT and pCT models are presented in Table 1. PCT model, could segment pCT images with high Dice 0.74-0.80 but failed to segment original CBCT (Dice 0.58-0.65) and CBCT model gave a moderate to low accuracy for segmenting weekly CBCTs as well as pCT. However, using sCBCT model, Dice range for segmenting both pCT and CBCTs were high 0.75-0.81 and 0.70-0.74, respectively, indicating unique potential of this physics based model that allows a cross-modality soft-tissue segmentation.

Table 1: Segmentation performance of the three models. HD represents 95[th] percentile Hausdorff distance between the ground truth and the segmentation contours.

| | Segmentation performance of sCBCT model | | | | | | | |
|---|---|---|---|---|---|---|---|---|
| | MSK | | External LIDC | | AAPM data | | MSK validation | |
| | Dice | HD (mm) | Dice | HD (mm) | Dice | HD (mm) | Dice | HD (mm) |
| CBCT | 0.74±0.04 | 3.9±1.1 | 0.73±0.04 | 3.9±1.0 | - | - | 0.70±0.08 | 5.6±2.0 |
| pCT | 0.81±0.04 | 2.6±0.6 | 0.78±0.01 | 2.5±0.4 | 0.75±0.07 | 3.3±1.4 | 0.77±0.09 | 3.9±2.2 |
| | Segmentation performance of pCT model | | | | | | | |
| | MSK | | External LIDC | | AAPM data | | MSK validation | |
| | Dice | HD (mm) | Dice | HD (mm) | Dice | HD (mm) | Dice | HD (mm) |
| CBCT | 0.65±0.1 | 6.7±4.0 | 0.60±0.08 | 13±7.2 | - | - | 0.58±0.1 | 10.8±8.1 |
| pCT | 0.80±0.04 | 2.6±0.6 | 0.76±0.02 | 2.8±0.8 | 0.74±0.04 | 4.5±3.9 | 0.77±0.09 | 3.4±1.7 |
| | Segmentation performance of CBCT model | | | | | | | |
| | MSK | | External LIDC | | AAPM data | | MSK validation | |
| | Dice | HD (mm) | Dice | HD (mm) | Dice | HD (mm) | Dice | HD (mm) |
| CBCT | 0.69±0.08 | 7.1±7.7 | 0.73±0.05 | 3.5±0.3 | - | - | 0.63±0.1 | 9.0±6.4 |
| pCT | 0.72±0.1 | 5.4±7.4 | 0.74±0.02 | 3.0±0.2 | 0.60±0.1 | 13.9±9.0 | 0.66±0.1 | 7.7±5.8 |

As a dosimetric evaluation of the segmentations, agreements between the ground-truth versus the sCBCT model esophagus segmentations evaluated using Bland-Altman method (Bland and Altman, 2007) showed absolute mean differences of 0.31 Gy (0.85% of population average mean dose) and 0.6 Gy (1.16% of population average $D5_{cc}$) for the mean dose and $D5_{cc}$ from the segmentations in CBCT images and 0.25 Gy (0.69% of population average) and 0.38 Gy (0.7% of population average) from the segmentations in pCT images of MSK validation cohort, respectively, indicating small differences. Moreover, dose volume histogram (DVH) of the ground-truth versus the sCBCT model esophagus segmentations were compared for two MSK validation cases on their pCT and CBCT images (Fig. 6A). Also, for the entire MSK validation cohort mean esophagus dose as well as $D5_{cc}$ calculated from the segmented contours showed good correlation with the ground-truth contours (Fig. 6B).

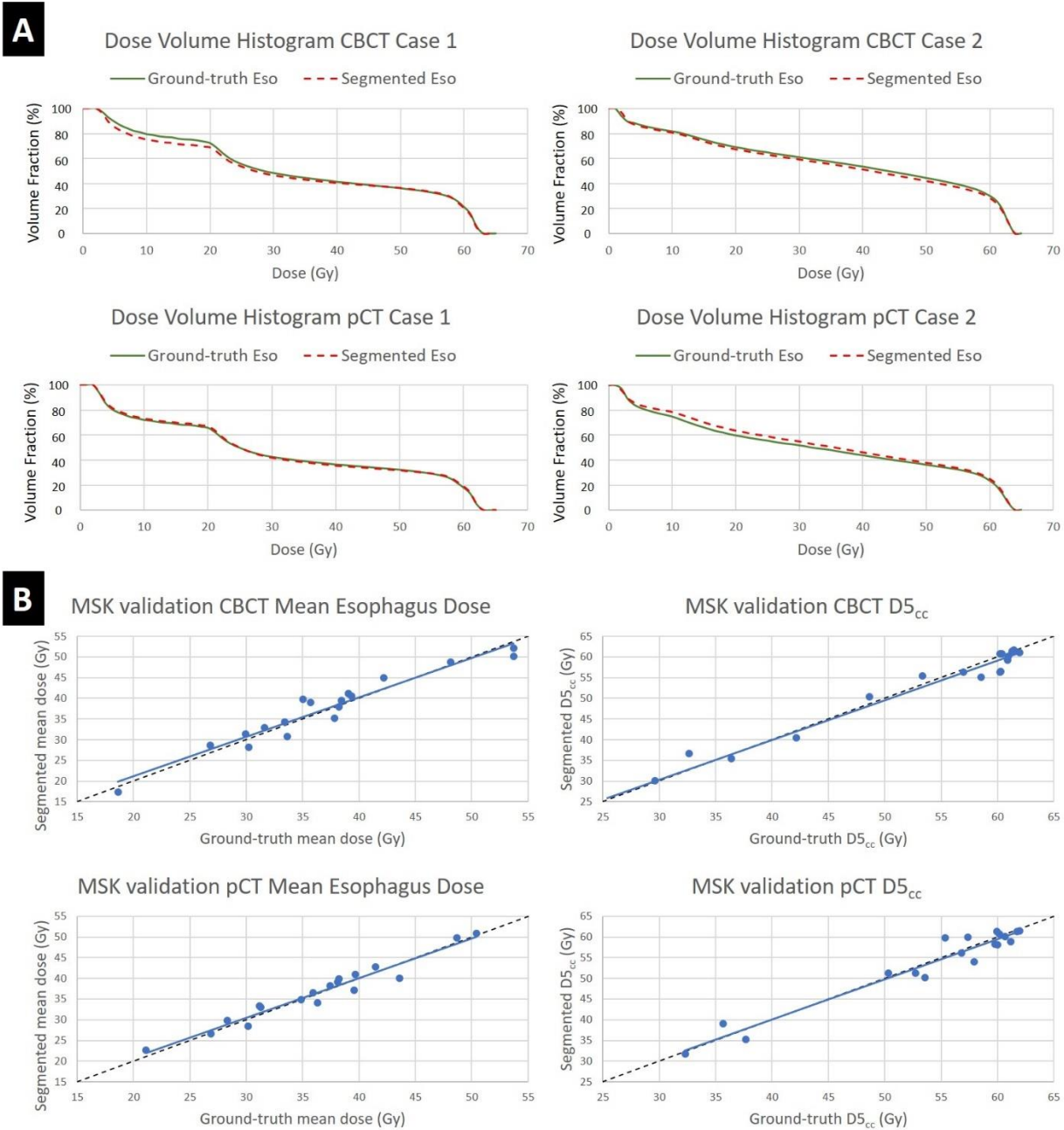

Fig. 6. A) Comparing dose volume histogram of ground-truth esophagus vs. segmentation for 2 typical MSK cases where their pCT and CBCTs esophagus were segmented using sCBCT model. B) Scatter plots showing correlations between mean esophagus dose and $D5_{cc}$ for the entire MSK validation cohort. Black dashed lines are identity lines.

## Discussion

Previous studies have highlighted segmentation differences between physicians or the same physician at different times on OARs and esophagus has had the lowest level of conformity among different OARs (30-32). This is mainly caused by esophagus's anatomic proximity to other mediastinal structures and similar HU value to nearby structures, which makes it difficult to delineate esophagus boundaries and its anatomical course. These also leads to high inter and intra-observer variability and institutional variability on segmentation. In contrast, by establishing a reference dataset and deformation registration, our 3D-UNet model could address this problem with an expressive consistency and overcome manual delineation error between observers. It is tedious and time consuming to manually define and contour the esophagus on each slice of the CT/CBCT images. The presented 3D-UNet segmentation could greatly improve delineation efficiency and reduce physician's workload. Meanwhile, robust auto-segmentation on wCBCTs can help physicians evaluate the real-time doses to the esophagus during the whole treatment process, which could provide reliable information to a timely adjustment of supportive treatment during radiotherapy.

In this novel study, we used pCT to generate multiple realistic (artifact-induced) sCBCT for 3D deep learning data augmentation. Artifacts/noises were extracted from CBCTs using PL-AHE method; an adaptive histogram equalization function that gives the users option to extract variations of contrasts from an image. Adaptive method computes several histograms, each corresponding to a distinct section of the image, and uses them to redistribute the lightness values of the image. It is therefore suitable for improving the local contrast and enhancing the definitions of edges in each region of an image. Even though, adaptive function operates on local regions of the image and transforms the local histograms to the local probability by fine-tuning the contrast effect parameters (α, β), it maintains the original histogram shape and does not affect the global histogram. This helps preserving local diagnostic characteristics of low-contrast soft tissues such as esophagus. In addition, PL-AHE does not cause some of the intensities to be overexposed and bright like standard histogram equalization. While on a global level it does not improve the contrast of large objects much, it does indeed make the more subtle low contrast soft-tissue features in the image more pronounced.

We validated our proposed model trained using sCBCT images on four different independent CBCTs and pCT data with the highest Dice of 0.74 and 0.81, respectively. The presented synthesis pipeline is significantly faster than the conventional Monte-Carlo based simulation process and results in large amounts of realistic augmented data for data-hungry 3D deep learning models. A 3D-UNet model trained on these realistic artifact-induced pCTs was robust and generalizable enough to segment esophagus on both weekly CBCTs and pCTs with high accuracy for longitudinal imaging studies. This model has the potential to segment any OAR on CBCT/pCT and therefore, can be used as a cross-modality segmentation tool for image guidance. Our results show that our semantic physics-based data augmentation spans the realistic noise/artifact spectrum across patient CBCT/pCT data and can generalize well across modalities, and can possibly segment (out-of-the-box) images reconstructed from latest CBCT reconstruction algorithms that lie on this spectrum (e.g. using Varian's new iterative cone beam CT iCBCT reconstruction method). Finally, it seems our artifacts-induced data augmentation results in "good inductive bias" (24) that forces the final deep learning model to become less sensitive to texture features and more biased towards global shape (similar to humans), in turn making it more generalizable to other modalities.

Recently, more attempts are being made to improve the CBCT image reconstruction algorithms by introducing for instance, the ability to utilize a linear accelerator's kV cone beam CT for treatment planning and dosimetry calculation using the Varian iterative cone beam CT (iCBCT) reconstruction method (Varian Medical Systems, Palo Alto, CA). This has improved intensity uniformity as well as better soft tissue visualization than standard reconstruction (Jarema and Aland, 2019). This work would open the way to taking advantage of this image modality, which is more easily available during treatment than time consuming MRI acquisition or labor-intensive methods that require expert guidance for assessing the RT response in real time rather than only for patient setup purposes.

In the future, we intend to additionally generate sCBCTs with motion artifacts by mimicking the motion from 4DCT images of these patients, essentially shifting sinogram projection to simulate motion artifact. Eventually, a deep learning model that can reproduce different types of artifacts e.g. scatter, motion, photon starvation, etc. for a better OAR segmentation to improve the accuracy of treatment and response analysis.